# MBE Growth of Cubic InN

Jörg Schörmann, Donat Josef As, and Klaus Lischka
Department of Physics, University of Paderborn, Warburger Strasse 100, Paderborn, 33095, Germany

## ABSTRACT

Cubic InN films were grown on top of a c-GaN buffer layer by rf-plasma assisted MBE at different growth temperatures. X-Ray diffraction investigations show that the c-InN layers consist of a nearly phase-pure zinc blende (cubic) structure with a small fraction of the wurtzite (hexagonal) phase grown on the (111) facets of the cubic layer. The content of hexagonal inclusions is decreasing with decreasing growth temperature. The full-width at half-maximum (FWHM) of c-InN (002) rocking curve is about 50 arcmin. Low temperature photoluminescence measurements reveal a band gap of about 0.61eV for cubic InN.

## INTRODUCTION

Among nitride semiconductors, InN is the least investigated of all and is expected to be a promising material for high frequency electronic devices [1,2]. The most important recent discovery about InN is that it has a much narrower band-gap than reported previously. For h-InN values between 0.6 eV and 0.7 eV are measured [3,4]. Group III-nitrides with cubic crystal structure are expected to have even lower band gaps and can be grown on substrates with cubic structure. However, the zincblende polytype is metastable and only a very narrow growth window is available for the process conditions [5]. The use of nearly lattice matched, free standing high quality 3C-SiC (001) substrates let to substantial improvements of the crystal quality of c-III-nitrides [6] and the absence of polarization fields [7] is advantageous for many device applications. So far little has been reported on the growth of cubic InN [8, 9].

## EXPERIMENT

Cubic InN films were grown on top of a c-GaN buffer layer (600 nm) by rf-plasma assisted molecular beam epitaxy (MBE). The c-GaN buffer layer was deposited on free standing 3C-SiC (001) substrates at growth temperatures of 720 °C. For the InN growth the temperature was reduced and varied in the range of 419°C to 490°C, respectively. InN growth was started under In rich conditions at an In-BEP of $6.8*10^{-8}$ mbar and was decreased to $3.1*10^{-8}$ mbar after two minutes of growth. The thicknesses of the InN layers were at least 130 nm and the growth was continuously monitored by reflection high energy electron diffraction (RHEED). Structural characterization was carried out by high resolution X-Ray diffraction (HRXRD). Photo-luminescence (PL) measurements at 10 K were performed using the 488 nm line of an $Ar^+$ laser and PL-signal was detected with an InAs photodiode.

## RESULTS AND DISCUSSION

HRXRD investigations were performed to determine the phase purity of our c-InN layers. All ω-2Θ-scans confirmed the formation of the cubic phase of InN. Bragg peaks observed at 35.8 °, 39.9 ° and 41.3 ° correspond to c-InN (002), c-GaN (002) and 3C-SiC (002), respectively. No additional reflexion of h-InN grown in (0002) direction was detected. The full width at half maximum (FWHM) of the (002) rocking curve was 48 arcmin. The lattice constant derived from the ω-2Θ-scan is 5.01 Å +/- 0.01 Å in good agreement to the values published [9].

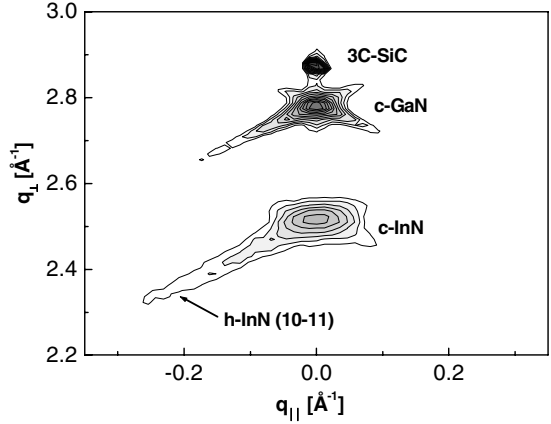

*Figure 1. Reciprocal space map of a c-InN layer. High intensity (002) Bragg reflexes of c-InN, c-GaN and the 3C-SiC substrate are observed.*

However, it is well known that hexagonal inclusions mainly grow on (111) facets and cannot be detected by ω-2Θ-scans. Therefore reciprocal space maps (RSM) of the GaN (002) Bragg-reflex were performed. The RSM along the (-110) azimuth is shown in Fig. 1. The growth temperature of this InN sample was 419 °C. From the intensity ratio of the cubic (002) reflex to the hexagonal (10-11) reflex we estimate 95 % cubic phase in this InN layer. RSM of the asymmetric GaN (-1-13) reflex show, that the lattice of our c-InN layers is fully relaxed with respect to the c-GaN buffer.

In Fig. 2 we plot the ratio the intensities of the h-(10-11) Bragg reflex and the c-(002) reflex versus the growth temperature of different InN layers. We observe a strong decrease of hexagonal inclusions with decreasing growth temperature up to a minimum value of 5 %. This supports the idea that with decreasing growth temperature the sticking coefficient of In is increasing, resulting in a higher density of cubic nuclei on the surface which reduce the formation of (111) facets.

Figure 3 shows the low temperature PL spectrum of an InN layer grown at 419 °C. The PL-spectrum show a peak at about 0.69 eV with a FWHM of 170 meV. We suppose that the broadening of the luminescence is due to the fact that the c-InN is degenerated. The lineshape of the luminescence (full curve in Fig. 3) was calculated using equ. 1 of Ref. 10. The low energy onset of the PL fitting curve gives the renormalized band-gap $E_g(n)$, which approaches the band-gap $E_g$ at vanishing free carrier concentration n= 0. We find a value of $E_g(n)$, which is close to 0.56 eV. Assuming a difference of about 50 meV between $E_g$ and $E_g(n)$ [10] we get a low temperature band-gap of c-InN close to 0.61 eV. By taking into account a temperature shift of about 50 meV between 10 K and 300 K the band-gap of c-InN is about 0.56 eV at room temperature.

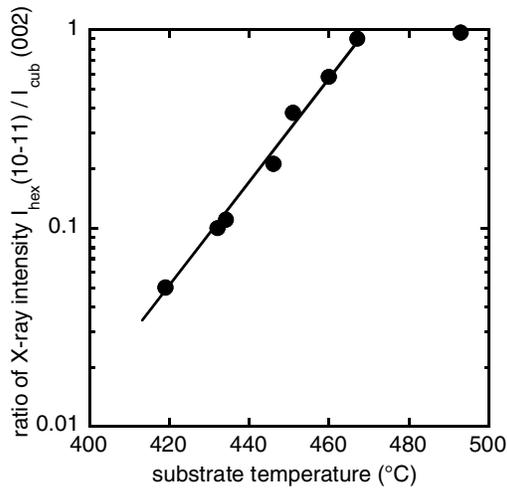

*Figure 2. Intensity ratio of the (10-11) reflex of hexagonal inclusions to the (002) Bragg reflex of cubic InN versus growth temperature of c-InN layers.*

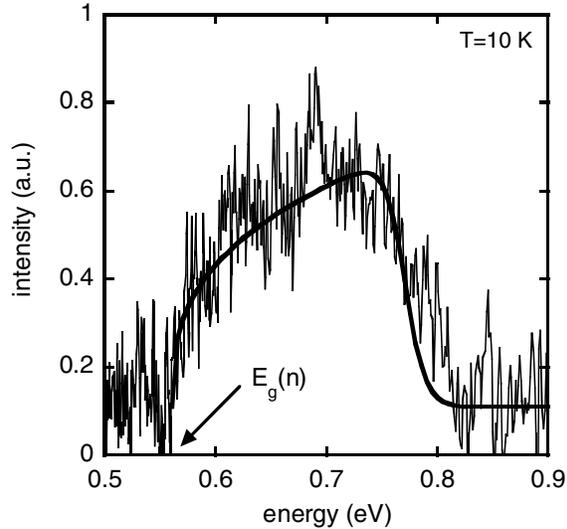

*Figure 3. Experimental and calculated photo-luminescence spectra of c-InN at 10 K.*

**CONCLUSIONS**

In summary cubic InN layers were grown on c-GaN buffer layers at different growth temperatures. The phase purity of c-InN layers increases with decreasing growth temperature. Cubic-InN layers with only 5 % hexagonal inclusions were grown at a substrate temperature of 419 °C. A minimum half width of the (002) XRD rocking curve of 48 arcmin and a lattice constant of 5.01 Å was measured. Low temperature PL measurements reveal a band gap of about 0.61 eV.

**ACKNOWLEDGMENTS**

We would like to thank Dr. H. Nagasawa and Dr. M. Abe from SiC Development Center, HOYA Corporation, for supplying the 3C-SiC substrates, and S.F. Li and W. Löffler for PL-measurements.